\documentclass[12pt]{article}

\usepackage{amsmath,amssymb,amsfonts,amsthm,url}
\usepackage{lmodern}
\usepackage{natbib}
\usepackage{xcolor,geometry,graphicx}
\def\init{\mathrm{init}}
\def\bA{\mathbf{A}}
\def\bH{\mathbf{H}}
\def\bb{\mathbf{b}}
\def\bbf{\mathbf{f}}
\def\bR{\mathbf{R}}
\def\bZ{\mathbf{Z}}

\def\ba{\mathbf{a}}

\def\cH{{\mathcal{H}}}

\def\cP{\mathcal{P}}
\def\bz{\mathbf{z}}
\def\bI{\mathbf{I}}
\def\bZ{\mathbf{Z}}

\def\bV{\mathbf{V}}
\def\bSigma{\boldsymbol{\Sigma}}
\def\bGamma{\boldsymbol{\Gamma}}
\def\bvphi{\boldsymbol{\varphi}}

\def\be{\mathbf{e}}
\def\bX{{\mathbf{X}}}

\def\bB{{\mathbf{B}}}

\def\cH{{\mathcal{H}}}

\def\ve{{\varepsilon}}
\def\balpha{{\boldsymbol{\alpha}}}
\def\bbeta{{\boldsymbol{\beta}}}
\def\bgamma{{\boldsymbol{\gamma}}}

\def\boldeta{\boldsymbol{\eta}}

\def\btheta{{\boldsymbol{\theta}}}
\def\R{{\mathbb{R}}}
\def\cC{\mathcal{C}}
\def\E{{\mathbb{E}}}
\def\P{{\mathbb{P}}}% \mathbb
\def\cP{{\mathcal{P}}}

\def\1{{\boldsymbol{1}}}
\def\0{{\boldsymbol{0}}}

\def\ext{\mathrm{ext}}
\def\var{{\mathrm{var}}}
\def\cov{{\mathrm{cov}}}
\def\eff{{\mathrm{eff}}}

\def\bmu{\boldsymbol{\mu}}

\newcommand*{\argmin}{\mathop{\arg\min}}

\def\ci{{\mathbin{\perp\mkern-10mu\perp}}}

\def\tf{\boldsymbol{f}}
\def\eff{{\mathrm{eff}}}
\def\os{{\mathrm{os}}}
\def\dd{{\mathrm{dd}}}
\def\aos{{\mathrm{aos}}}

\def\diag{\mathrm{diag}}

\newtheorem{theorem}{Theorem}%[section]
\newtheorem{proposition}{Proposition}

\newtheorem{remark}{Remark}
\newtheorem{example}{Example}

\begin{document}

\def\spacingset#1{\renewcommand{\baselinestretch}%
{#1}\small\normalsize} \spacingset{1}

%%%%%%%%%%%%%%%%%%%%%%%%%%%%%%%%%%%%%%%%%%%%%%%%%%%%%%%%%%%%%%%%%%%%%%%%%%%%%%

  \title{\bf Efficient Semiparametric Inference for Distributed Data with Blockwise Missingness}
  \author{Jingyue Huang\\
    % Department of YYY, University of XXX\\
    and \\
    Huiyuan Wang \\and \\
    Yuqing Lei\\ and \\
    % Department of ZZZ, University of WWW\\ 
    Yong Chen\thanks{Corresponding author: \texttt{ychen123@pennmedicine.upenn.edu}
    % The authors gratefully acknowledge \textit{please remember to list all relevant funding sources in the version that gives all author information}
    }\hspace{.2cm} \\
    {\em \small Department of Biostatistics, Epidemiology, and Informatics}\\ {\em \small Perelman School of Medicine, University of Pennsylvania}\\ {\em \small Philadelphia, PA, USA}
    % \\ 
    % The Center for Health AI and Synthesis of Evidence (CHASE),\\ University of Pennsylvania,\\ Philadelphia, PA, USA
    }
  \maketitle

\bigskip
\begin{abstract}
We consider statistical inference for a finite-dimensional parameter in a regular semiparametric model under a distributed setting with blockwise missingness, where entire blocks of variables are unavailable at certain sites and sharing individual-level data is not allowed. 
To improve efficiency of the internal study, we propose a class of augmented one-step estimators that incorporate information from external sites through ``transfer functions.'' 
The proposed approach has several advantages. 
First, it is communication-efficient, requiring only one-round communication of summary-level statistics. 
Second, it satisfies a do-no-harm property in the sense that the augmented estimator is no less efficient than the original one based solely on the internal data. 
Third, it is statistically optimal, achieving the semiparametric efficiency bound when the transfer function is appropriately estimated from data. 
Finally, it is scalable, remaining asymptotically normal even when the number of external sites and the data dimension grow exponentially with the internal sample size. 
Simulation studies 
%and a real-world data analysis 
confirm both the statistical efficiency and computational feasibility of our method in distributed settings. 
\end{abstract}

\noindent%
\vfill

\newpage
\spacingset{1.8} % DON'T change the spacing!

\section{Introduction}\label{sec-intro}
Multi-center research has attracted considerable attention for its potential to enhance the statistical efficiency and power of single-site studies through collaborative efforts \citep{li2024centralized}.
However, a persistent challenge in such collaborative research is blockwise missingness, a systematic issue where some sites consistently collect only specific, and potentially overlapping, subsets of variables. 
This problem often arises from differences in study designs, data collection protocols, or resource constraints \citep{madden2016missing}. 
For example, in the Alzheimer’s Disease Neuroimaging Initiative, certain sites lack measurements for specific modalities, such as advanced imaging, due to equipment limitations or site-specific decisions \citep{mueller2005alzheimer}. 
Such blockwise missingness significantly restricts the utility of external data in integrated analyses, thereby limiting their statistical and scientific potential.

To address blockwise missingness, several methods have been developed that leverage specific structures in the working model or the underlying data distribution. 
For instance, the multiple blockwise imputation approach exploits the linear structure of the working model to construct estimating equations that integrate imputations from both complete and incomplete observations \citep{xue2021integrating}. 
Building on a geometric interpretation of linear regression coefficients, \citet{song2024semi} proposed a regularized empirical risk minimization framework that avoids imputation. 
Also for linear models, \citet{yu2020optimal} emphasized the importance of the complete-data covariance matrix, where they proposed to aggregate intra-block and cross-block covariance matrices in a linear form to obtain more accurate estimates. 
Another related approach, modular regression \citep{Jin2023}, utilizes the conditional independence assumption across missing blocks given the always-observed block to improve the estimation of the complete-data covariance matrix.
While these methods offer practical solutions for specific problems, such as (generalized) linear regression, they often lack theoretical guarantees on their optimality and are narrowly tailored to specific statistical tasks, thereby failing to meet the growing demands of multi-center research.

The challenge becomes even more pronounced in distributed settings, where sharing individual-level data is restricted by privacy regulations such as the Health Insurance Portability and Accountability Act \citep{annas2003hipaa}. Distributed networks that enable the exchange of summary-level information provide an attractive alternative, allowing efficient inference without direct access to individual data. Distributed learning algorithms have been proposed in a wide range of contexts, including high-dimensional linear regression, (semi)parametric $M$-estimation, and Bayesian inference \citep[see, e.g.,][and references therein]{Battey2018,Jordan2019,Duan2022,fan2023communication}.
However, extending these distributed methods to accommodate blockwise missing data remains unclear, as does developing distributed counterparts of the statistical methods for blockwise missingness discussed above. An additional practical consideration is the minimization of communication rounds across local sites. In large-scale distributed networks such as the Observational Health Data Sciences and Informatics initiative \citep{hripcsak2015observational}, coordinating iterative updates among international collaborators can be prohibitively time-consuming \citep{li2024centralized}. Addressing these challenges calls for distributed algorithms that are both communication-efficient and statistically robust to blockwise missingness.

In this paper, we introduce a meta-algorithm, termed augmented one-step (AOS) estimation, for the setting where an internal site contains fully observed data and multiple external sites observe only specific subsets of variables. 
Unlike traditional methods tailored to specific tasks, this meta-algorithm provides a flexible framework that integrates two key components: an initial one-step estimator \citep{pfanzagl1982lecture} using fully observed internal data, and an augmentation step leveraging transfer functions that summarize useful information from external sites. 
The AOS framework maintains the broad applicability of traditional one-step estimators while offering several additional advantages:
\begin{enumerate}
\item {\it Communication efficiency}: It requires only a single round of transmitting summary-level statistics from local sites.
\item{\it Do-no-harm property}: It is at least as efficient as the estimator obtained from the internal study, irrespective of the choice of transfer functions, correlation strength among blocks of variables, and the missing patterns in external sites, which is desirable in domains such as AI in healthcare \citep{goldberg2024no}.
\item {\it Statistical optimality}: When the optimal transfer function is estimated under a correctly specified working model, the AOS estimator achieves the semiparametric efficiency bound.
\item {\it Scalability}: The method remains effective as the number of external sites or the data dimension increases, making it suitable for large-scale, multi-center studies.
\end{enumerate}
It is worth noting that identifying and estimating the optimal transfer function involves solving estimating equations subject to multiple, overlapping conditional moment restrictions. While in this paper we address the problem using the kernel trick, one could also borrow general techniques from semiparametric estimation and econometrics \citep{ai2003efficient, carrasco2007linear}. In this sense, our work highlights a novel application setting for such methods.

\subsection{Related Work}\label{sec:related_work}
To improve the efficiency of single-site studies, a classical strategy is to fuse internal data with external summary statistics. For example, \citet{chatterjee2016constrained} proposed constrained maximum likelihood estimators that incorporate external moments as constraints, and \citet{zhang2020generalized} extended this framework to account for variability in external summaries and potential model misspecification. More recently, \citet{hu2022paradoxes} developed a semiparametric, influence-function–based method that alleviates the computational cost of empirical likelihood and accommodates mismatches between internal and external target parameters. In these approaches, the summary statistics correspond to predefined low-dimensional parameters, and it is unclear how to extend them to blockwise missingness. Furthermore, these methods do not address which parameters are optimal for data fusion and how such parameters should be estimated.

Prediction-powered inference (PPI) \citep{angelopoulos2023prediction} goes beyond transferring finite dimensional parameters. This line of work leverages predictions from pretrained models to reduce estimator variance in semi-supervised settings, where the pretrained models can be arbitrarily complex, and even overparametrized. Interestingly, by choosing a specific transfer function to be defined in Section \ref{sec:aug_given_tf}, the proposed estimator matches a variant of PPI estimator introduced in \citet{xu2025unified}, which addresses the negative transfer issue in the vanilla PPI estimator. This ``do-no-harm'' property is maintained by our proposed method when dealing with distributed blockwise missing data.

Semi-supervised learning, where only the outcome is missing, can be regarded as a special case of blockwise missingness; see Remark~\ref{remark:ssi}. Related methods \citep[e.g.,][]{zhang2019semi,cannings2022correlation} used the control variate technique to enhance the efficiency of estimators based only on labeled data. Our work can be viewed as a generalization of their methods to much more complex blockwise-missingness scenarios, in particular with distributed data. We establish the semiparametric efficiency bound for distributed data with blockwise missingness, and further show that our proposed method attains the efficiency bound under certain assumptions.

% \begin{longtable}[]{@{}lllll@{}}
% \caption{D-optimality values for design X under five different
% scenarios.}\label{tbl-one}\tabularnewline
% \toprule\noalign{}
% one & two & three & four & five \\
% \midrule\noalign{}
% \endfirsthead
% \toprule\noalign{}
% one & two & three & four & five \\
% \midrule\noalign{}
% \endhead
% \bottomrule\noalign{}
% \endlastfoot
% 1.23 & 3.45 & 5.00 & 1.21 & 3.41 \\
% 1.23 & 3.45 & 5.00 & 1.21 & 3.42 \\
% 1.23 & 3.45 & 5.00 & 1.21 & 3.43 \\
% \end{longtable}

\section{Preliminaries}\label{sec:pre}
Consider a system with one \emph{internal site} and $J$ \emph{external sites}, each possessing a local sample of potentially different sizes.
Denote the sample size of the internal site by $n_0$ and that of the $j$th external site by $n_j$. 
For notational simplicity, let $S_j$ be the index set of samples from the $j$th site, $j=0,1,\dots,J$; for example, $S_0=\{1,\dots,n\}$ and $S_j=\{1+\sum_{k=0}^{j-1}n_k,\dots,\sum_{k=0}^{j}n_k\}$ for $j\geq 1$, with the convention that $n_0=n$. We allow for the extremely heterogeneous sample sizes across sites, i.e., $n/n_j\to0$ as $n\to\infty$, which violates the standard positivity assumption in missing data literature. 
% Our primary focus is to improve estimation efficiency in distributed settings with blockwise missingness. 

Let $\cP$ represent a class of data distributions defined on the sample space $\Omega\subset\R^q$, which is referred to as the model.  The internal site provides a sample $\{\bZ_i:i\in S_0\}$, where each $\bZ_i$ is an independent realization of the random vector $\bZ=(Z_{1},\dots,Z_{q})^\top\sim P_0\in\cP$. 
The $d$-dimensional target parameter $\btheta^*$ is defined with respect to the distribution $P_0$.

External studies, on the other hand, may have different objectives and observe only subsets of $\bZ$. 
For the $j$th ($j=1,\dots,J$) external site, let $B_j\subset\{1,\dots,p\}$ index the observed variables. 
Accordingly, the $j$th external site observes $n_j$ independent copies of the random vector $\bZ_{B_j}=(Z_b)_{b\in B_j}^\top\sim Q_j$, which we represent as $\{\bZ_{i,B_j}:i\in S_j\}$.  
The missingness structure may vary across external sites but is consistent within each site, and we allow that $B_j$ may be an arbitrary subset of $\{1,\dots,p\}$. This setting reflects practical scenarios where different sites may record different subsets of variables, which cannot be fully captured by some assumptions commonly used in the missing data literature such as monotone coarsening \citep{tsiatis2006semiparametric}.
Moreover, we allow for heterogeneous data distributions across sites to some extent. In particular, the distribution $Q_j$ at the $j$th external site need not correspond to any marginal distribution of the internal site distribution $P_0$. See the discussion following Theorem \ref{thm:main_nonsplit} for details.

The true value of the target parameter $\btheta^*$ can be viewed as the value of the map $P\mapsto\btheta(P)$ evaluated at the internal data distribution, that is, $\btheta^*=\btheta(P_0)$. 
Throughout the paper, we assume that the map $\btheta(\cdot)$ has an influence function $\bvphi(\bZ;\btheta,\boldeta)$ under the model $\cP$, where $\boldeta$ represents the nuisance parameter. 
% associated with the target parameter. 
% For technical simplicity, we further assume that $\eta\in\cG$ and $\cG$ is a known linear subset of $L_2(P)=\{g:\E_{P}[\{g(Z)\}^2]<\infty\}$ or simply $L_2(P)$ itself. 
Detailed conditions for the existence of the influence function are provided in Section S.2 in the Supplementary Material.

Consider a scenario where initial estimators $(\widehat\btheta^\init,\widehat\boldeta^\init)$ for the target and nuisance parameters are obtained from the internal study. 
Under certain conditions \citep{pfanzagl1982lecture, newey1994large}, the one-step estimator, defined as
\begin{equation}\label{eq:os}
\widehat\btheta^\os=\widehat\btheta^\init+n^{-1}\sum_{i\in S_0}\bvphi(\bZ_i;\widehat\btheta^\init,\widehat\boldeta^\init),
\end{equation}
is regular and asymptotically linear in the sense that 
\[
n^{1/2}\bigl(\widehat\btheta^\os-\btheta^*\bigr)=n^{-1/2}\sum_{i\in S_0}\bvphi(\bZ_i;\btheta^*,\boldeta^*)+o_P(1),
\]
where $\boldeta^*$ is the true value of the nuisance parameter.
Thus, the one-step estimator is asymptotically normal with variance $\var_{P_0}\{\bvphi(\bZ;\btheta^*,\boldeta^*)\}$. The asymptotic variance can be consistently estimated by the sample variance of $\bvphi(\bZ_i;\widehat\btheta^\init,\widehat\boldeta^\init)$, enabling valid statistical inference.
To clarify these concepts, we provide a concrete example below.

\setcounter{example}{0}
\begin{example}[Mean of potential outcomes] 
Let $\bX$ denote the confounders, $A$ the binary treatment assignment, and $Y$ the outcome of interest.  
Under the potential outcome framework \citep{rubin1976inference}, let $Y^a$ represent the potential outcome if a subject receives treatment $A=a\in\{0,1\}$. 
The model $\cP$ consists of data distributions $P$ for $\bZ=(\bX^\top,A,Y)^\top$ satisfying
\[
Y=AY^1+(1-A)Y^0,\quad A\ci(Y^1,Y^0)\mid \bX,\quad\P_{P_X}\{\varepsilon\le\P_P(A=1\mid \bX)\le1-\varepsilon\}=1
\]
for some constant $\ve\in(0,1)$ with $\bX\sim P_X$. 
The target parameter is the mean of $Y^a$, $$\theta(P)=\E_P(Y^a)=\E_{P_X}\{g_a(\bX;P)\}=\E_P\bigl\{I(A=a)Y/\pi_a(\bX)\bigr\},$$
where $g_a(\bX)=\E_P(Y\mid \bX,A=a)$ and $\pi_a(\bX)=\P_P(A=a\mid \bX)$ are nuisance parameters. 
The efficient influence function under $\cP$ is \citep{hahn1998role}
\[
\bvphi(\bZ;\theta,g_a,\pi_a)=\frac{I(A=a)}{\pi_a(\bX)}\{Y-g_a(\bX)\}+g_a(\bX)-\theta.
\]
Using the initial estimator $\widehat{g}_a^\init(\bX)$ or $\widehat\pi^\init_a(\bX)$ from the internal sample, we compute the initial estimator for the target parameter $\widehat\theta^\init=n^{-1}\sum_{i\in S_0}\widehat{g}^\init_a(\bX_i)$ or $\widehat\theta^\init=n^{-1}\sum_{i\in S_0}I(A_i=a)Y_i/\widehat{\pi}^\init_a(\bX_i)$. 
The one-step estimator is then calculated using equation \eqref{eq:os}.
\end{example}

% The one-step estimator is necessary when the initial estimator for the target parameter is not root-$n$ consistent \citep{kennedy2022semiparametric,chernozhukov2018double}. 
% Section \ref{sec:pd_if_vme} in the Supplementary Material demonstrates that the first-order bias of the initial estimator can be expressed as $-\int_\Omega\bvphi(z;\widehat\theta^\init,\widehat\eta^\init)dP(z)$. 
% The one-step estimator in \eqref{eq:os} corrects this bias by using its estimate $n^{-1}\sum_{i\in S_0}\bvphi(Z_i;\widehat\theta^\init,\widehat\eta^\init)$. 
% In principle, any estimator of the first-order bias can be used to construct the one-step estimator, but we need to choose the one that yields the minimum asymptotic variance. 

For simplicity, write $\widehat\bvphi(\bz)=\bvphi(\bz;\widehat\btheta^\init,\widehat\boldeta^\init)$ and $\bvphi(\bz)=\bvphi(\bz;\btheta^*,\boldeta^*)$. It is noteworthy that the asymptotic variance of the one-step estimator, $\var_{P_0}(\bvphi)$, is the semiparametric efficiency bound when $\bvphi(\bZ)$ is the efficient influence function and other regularity conditions hold \citep{bickel1993efficient}. 
In this case, given only the internal sample, no estimator can achieve a smaller mean squared error than the one-step estimator in the local minimax sense (Corollary~2.6 in \citealt{vander2002}). 

While the one-step estimator remains asymptotically efficient when using only the internal study, its variance can be large when the internal sample size is limited—a common scenario in real-world analyses. To address this issue, in the next section we develop a principled method that effectively leverages them to enhance the efficiency of internal study estimation, although the external studies in our setting exhibit blockwise missingness. To highlight the efficiency gain, for the remainder of the paper we let $\bvphi(\bZ)$ denote the efficient influence function for $\btheta(P)$ under model~$\cP$.

% We first introduce some notation throughout the rest of the paper.  Given an independent and identically distributed (i.i.d.) sample $Z_1,\dots,Z_n$, and two functions $f,g:\Omega\to\R^d$, we write $\widehat\E_n(f)=n^{-1}\sum_{i\in S_0}f(Z_i)$ as the empirical expectation of $f$ and $\widehat\var(f,g)=n^{-1}\sum_{i\in S_0}\{f(Z_i)-\widehat\E_n(f)\}\{g(Z_i)-\widehat\E_n(g)\}^\top$ the empirical covariance matrix. On the population level, we write $\E_P(f)=\int_{\Omega}f(z)dP(z)$ the expectation of $f(Z)$ under $Z\sim P$ and $\var_{P_0}(f,g)=\E_P[f(Z)\{g(Z)\}^\top]-\E_P(f)\{\E_P(g)\}^\top$ the covariance matrix.  For brevity, we rewrite $\widehat\var(f,f)$ as $\widehat\var(f)$, $\var(f,f)$ as $\var(f)$, and $\bvphi(z;\widehat\theta^\init,\widehat\eta^\init)$ as $\widehat\bvphi(z)$.

% \begin{example}[Generalized linear models]
% Let $X=(X_1,\dots,X_d)^\top$ be a vector of covariates and $Y$ the binary outcome of interest. Our model $\cP$ is chosen as the set of all data distribution for $Z=(X^\top,Y)^\top$ satisfying
% \[
% X\sim P_X,\quad \P_P(Y=1\mid X)=g(X_1\beta_1+\dots+X_d\beta_d)
% \]
% with $g(t)=e^t/(1+e^t)$. The parameter of interest is the regression coefficient of the $j$th predictor, $\theta(P)=e_j^\top\beta$, where $e_j$ denotes the $j$th column of the $d\times d$ identity matrix. Let $\eta=\beta$ be the nuisance parameter, and thus $m(Z,\eta)=e_j^\top\eta$ identifies $\theta(P)$.
% \end{example}

\section{Augmented one-step estimator}

\subsection{Augmentation with a given transfer function}\label{sec:aug_given_tf}
Recall that our setting involves one internal site and \(J\) external sites, where the internal site observes complete data \(\bZ \sim P_0\) and each external site \(j\) observes only a subset \(\bZ_{B_j} \sim Q_j\). The missingness pattern is blockwise, potentially with overlapping observed subsets across sites, and the external-site distributions \(Q_j\) may differ from \(P_0\) beyond simple marginalization. This heterogeneity, combined with non-overlapping variables, poses a challenge: directly leveraging external data to improve the estimation efficiency of the target parameter \(\btheta^*\) is nontrivial.  

We propose to use the control-variate technique \citep{nelson1990control} to construct an estimator that is more efficient than the one-step estimator with the efficient influence function $\bvphi(\cdot)$. The control-variate technique is popular in estimating the expected value of a function, particularly in the context of Monte Carlo integration. 
In this method, a control variate is a random variable that correlates with the target integrand and has a known expected value. 
% In our case, the first-order bias $\int_\Omega\widehat\bvphi(Z)dP(z)$ is exactly the integral of the efficient influence function $\widehat\bvphi(z)$. 
Suppose that $\tf(\bZ)$ is a vector-valued function with a known expectation $\E_P\{\tf(\bZ)\}$ and invertible variance matrix $\var(\tf)$. 
Consider an augmented influence function $\bvphi_f(\bZ)\equiv\bvphi(\bZ)-\cov(\bvphi,\tf)\{\var(\tf)\}^{-1}\{\tf(\bZ)-\E(\tf)\}$.
This function retains the same expectation as $\bvphi(\bZ)$, but its variance satisfies
\[
\var(\bvphi_f)=\var(\bvphi)-\cov(\bvphi,\tf)\{\var(\tf)\}^{-1}\cov(\tf,\bvphi)\preceq\var(\bvphi),
\]
where for two square matrices $\bA_1$ and $\bA_2$, $\bA_1\preceq\bA_2$ if $\bA_2-\bA_1$ is positive semi-definite. 
Thus, the sample mean $n^{-1}\sum_{i\in S_0}\bvphi_f(\bZ_i)$ exhibits a smaller asymptotic variance than the original mean $n^{-1}\sum_{i\in S_0}\bvphi(\bZ_i)$. 
In fact, $\bvphi_f$ is the efficient influence function for $\btheta(P)$ under the additional restriction that $\E_P\{\tf(\bZ)\}$ is known; see Example 3 in Section 3.2 of \citet{bickel1993efficient}. 
If $\E_P\{\tf(\bZ)\}$ were known in advance, we could replace $n^{-1}\sum_{i\in S_0}\widehat\bvphi(\bZ_i)$ in  \eqref{eq:os} with $n^{-1}\sum_{i\in S_0}\widehat\bvphi_f(\bZ_i)$, where $\widehat\bvphi_f(\bZ)=\widehat\bvphi(\bZ)-\widehat\cov(\widehat\bvphi,\tf)\{\widehat\var(\tf)\}^{-1}\{\tf(\bZ)-\E(\tf)\}$ denotes an estimate of $\bvphi_f(\bZ)$ with $\widehat\cov(
\widehat\bvphi,\tf)$ and $\widehat\var(\tf)$ being the sample covariance and variance at the internal site, respectively.
% There is a correspondence between an estimator and its influence function.

In our distributed setting with blockwise missingness, however, the expectation of $\tf(\bZ)$ is generally unknown and must be estimated. 
We achieve this by leveraging external sites. 
Moreover, each external site may have different blockwise missingness patterns, and thus site-specific functions are required. 
Let $\tf_j(\cdot):\bZ_{B_j}\mapsto \tf_j(\bZ_{B_j})\in\R^{p_j}$ denote the function for the $j$th external site.
We term $\tf_j(\cdot)$ the \emph{transfer function}, emphasizing that the sample mean of $\tf_j(\bZ_{B_j})$ at the $j$th external site summarizes the information to be transferred to the internal site. Define $\bbf(\bZ)=\{\tf_1(\bZ_{B_1})^\top, \dots, \tf_J(\bZ_{B_J})^\top\}^\top$ as the aggregated transfer function.
Following the conventional control-variate technique, and by replacing the expectation of $\bbf(\bZ)$ with its estimate obtained from external sites, the augmented influence function is 
\[
\bvphi_{f}(\bZ;\bbeta)\equiv\bvphi_{f}(\bZ;\btheta^*,\boldeta^*,\bbeta)=\bvphi(\bZ;\btheta^*,\boldeta^*)-\bbeta^\top\{\bbf(\bZ)-\widetilde{\bmu}_\ext\},
\]
where $\widetilde{\bmu}_{\ext}=(\widetilde{\bmu}_{\ext,1}^\top, \dots, \widetilde{\bmu}_{\ext,J}^\top)^\top$ with $\widetilde{\bmu}_{\ext,j}=n_j^{-1}\sum_{i\in S_j} \tf_j(\bZ_{i,B_j})$ representing the estimate for the expected value of $\tf_j(\bZ_{B_j})$, and ${\bbeta}\in\mathbb{R}^{p\times d}$ is a nuisance parameter with $p=\sum_{j=1}^J p_j$ which we refer to as the transfer coefficients.

The function $\bvphi_{f}(\bZ;\bbeta)$ shares the same expectation with $\bvphi(\bZ)$, but its variance depends on the transfer coefficients $\bbeta$.
We then choose the optimal transfer coefficients by minimizing the variance $\var_{P_0}\{n^{-1}\sum_{i\in S_0}\bvphi_{f}(\bZ_i;\bbeta)\}$.
This variance should be computed with respect to the randomness of $\{\bZ_i:i\in S_0\}$ and $\widetilde\bmu_\ext$, as the randomness of $\bbf(\bZ_i)$ arises from $\bZ_i$.
Given the independence of the internal and external samples, straightforward calculations yield
\begin{equation}
\bbeta^*=\argmin_{\bbeta\in\mathbb{R}^{p\times d}} \bbeta^\top\bigl\{\var_{P_0}({\bbf})+\bSigma_\ext\bigr\}\bbeta -2\bbeta^\top\cov_{P_0}(\bbf,\bvphi),
\label{eq:beta*}
\end{equation}
where $\bSigma_\ext=\diag(\rho_1\bSigma_{\ext,1},\dots, \rho_J\bSigma_{\ext,J})$ with $\rho_j=\lim_{n\to\infty}n/n_j$ being the ratio of internal and external sample sizes and $\bSigma_{\ext,j}=\var_{Q_j}\{\tf_j(\bZ_{B_j})\}$ denoting the variance of $\tf_j(\bZ_{B_j})$ from the $j$th external site. 
% \begin{remark}

% \end{remark}
To estimate the optimal transfer coefficients $\bbeta^*$, one may solve the (regularized) empirical counterpart of \eqref{eq:beta*}:
\begin{equation}\label{eq:betahat}
\widehat{\bbeta}=\argmin_{\bbeta\in\mathbb{R}^{p\times d}} \bbeta^\top\bigl\{\widehat{\var}({\bbf})+\widetilde{\bSigma}_\ext\bigr\}\bbeta -2\bbeta^\top\widehat{\cov} \bigl(\bbf, \widehat\bvphi\bigr)+\lambda J(\bbeta), 
\end{equation}
where $\widehat{\var}({\bbf})$ and $\widehat{\cov} \bigl(\bbf, \widehat\bvphi\bigr)$ are the sample variance and covariance calculated from the internal sample, $\widetilde{\bSigma}_\ext=\diag(\rho_1\widetilde{\bSigma}_{\ext,1},\dots, \rho_J\widetilde{\bSigma}_{\ext,J})$ with $\widetilde{\bSigma}_{\ext,j}$ representing the sample variance from the $j$th external site, $J(\bbeta)$ denotes the penalty term, and $\lambda$ is the tuning parameter.
The penalty term $J(\bbeta)$ can be discarded by setting $\lambda=0$. However, it is recommended to keep it when the dimension of the aggregated transfer function exceeds the internal sample size, or when prior information suggests that some external sites may not provide useful information for parameter estimation. We emphasize that only a consistent estimator of $\bbeta^*$ is required to achieve the efficiency gain of the proposed method, and any other consistent alternatives of \eqref{eq:betahat} are admissible; see Theorem~\ref{thm:main_nonsplit} for details.

Finally, the augmented one-step (AOS) estimator, incorporating external information, is
\begin{equation}\label{eq:aos}
\widehat{\btheta}^{\aos}=\widehat\btheta^\init+n^{-1}\sum_{i\in S_0}\bigl\{\widehat\bvphi(\bZ_i)-\widehat{\bbeta}^\top(\bbf_i-\widetilde{\bmu}_\ext)\bigr\},
\end{equation} 
where for simplicity, we write $\bbf_i=\bbf(\bZ_i)$ as the function evaluation on the $i$th internal data point.
% and it is more efficient than the OS estimator.
Notably, the AOS estimator requires only summary-level statistics $(n_j,\widetilde{\bmu}_{\ext,j},\widetilde{\bSigma}_{\ext,j})$ from external sites, making it communication-efficient and practical for integrating external data. 

To clarify the relationship between our proposed method and existing work, we provide two remarks below.

\begin{remark}[Connections to the Semi-supervised Inference]\label{remark:ssi}
Semisupervised learning is a special case of learning with blockwise missing data, involving the following setup: $\bZ=(Y,\bX^\top)^\top$, the internal study observes labeled data $\{(Y_i,\bX_i)\}_{i=1}^n$, while the external study observes only covariates $\{\bX_i\}_{i=n+1}^{n+n_1}$. 
When the target parameter is the expectation of $Y$, the influence function simplifies to $\widehat\bvphi(\bZ)=Y-\widehat\btheta^\init$, where $\widehat\btheta^\init=n^{-1}\sum_{i\in S_0}Y_i$. 
Setting $\bbf(\bZ)=\bX$, we obtain $\widetilde\bmu_\ext=n_1^{-1}\sum_{i=n+1}^{n+n_1}\bX_i$, which is the sample mean of $\bX$ from the external site. 
A consistent estimator of $\bbeta^*$ in \eqref{eq:beta*} takes the explicit form $\widehat\bbeta=n_1\widehat\bgamma/(n_1+n)$, where $\widehat\bgamma=\{\widehat\var(\bX)\}^{-1}\widehat\cov(\bX,Y)$.
%  and $\widehat\var(\bX)$ is the sample variance of $\bX$ on the labeled data. 
Consequently, the augmented one-step estimator becomes 
\[
n^{-1}\sum_{i\in S_0}\{Y_i-\widehat\bbeta^\top(\bX_i-\widetilde\bmu_\ext)\}=n^{-1}\sum_{i\in S_0}Y_i-\widehat\bgamma^\top\biggl\{n^{-1}\sum_{i\in S_0}\bX_i-(n+n_1)^{-1}\sum_{i=1}^{n+n_1}\bX_i\biggr\},
\]
which coincides with the semi-supervised least squares (SSLS) estimator proposed by \citet{zhang2019semi}. 
Thus, our proposed augmented one-step estimator generalizes the SSLS estimator to accommodate more complex blockwise missingness in distributed settings and applies to any estimand that has an associated influence function.
\end{remark}

\begin{remark}[Connections to Prediction-Powered Inference]\label{remark:ppi}
Prediction-powered inference (PPI) and its variants \citep{angelopoulos2023prediction} improve $M$-estimation efficiency in semi-supervised settings by leveraging predictions from pretrained models. 
Under the setup in Remark~\ref{remark:ssi}, let $\btheta^*\in\R^d$ minimize $\E_{P_0} \{\ell(\btheta; Y, \bX)\}$ for a general loss function $\ell(\cdot;\cdot)$, and let $\widehat\btheta^{\init}=\argmin_{\btheta}n^{-1}\sum_{i=1}^n\ell(\btheta;Y_i,\bX_i)$ be the supervised estimator, whose influence function is $\bvphi(\bz;\btheta)=[\E_{P_0}\{\nabla^2_{\btheta}\ell(\btheta^*;Y,\bX)\}]^{-1}\nabla_{\btheta}\ell(\btheta;Y,\bX)$ under regularity conditions \citep{newey1994asymptotic}. 

With a pretrained model $\widehat{h}(\cdot)$, we can choose the transfer function $\bbf(\bZ)=\nabla_{\btheta}\ell(\widehat\btheta^{\init};\widehat{h}(\bX),\bX)$, leading to $\widehat\bbeta=n_1\widehat\bgamma/(n+n_1)$, where $\widehat\bgamma=[\widehat\var_{\mathrm{full}}\{\bbf(\bX)\}]^{-1}\widehat\cov\{\bbf(\bX),\bvphi(\bZ,\widehat\btheta^{\init})\}$ and $\widehat\var_{\mathrm{full}}\{\bbf(\bX)\}$ is the full sample variance of the transfer function. 
The augmented estimator
\[
\widehat\btheta^{\init} - \widehat\bgamma^\top\left\{ n^{-1} \sum_{i = 1}^n \bbf(\bX_i) -(n+n_1)^{-1} \sum_{i=1}^{n+n_1} \bbf(\bX_i) \right\}
\]
coincides with the PPI-type estimator of \citet{xu2025unified}, which is \emph{safe} in that it is never less efficient than the supervised estimator, regardless of model quality or the proportion of unlabeled data. 
Our augmented one-step estimator recovers this PPI-type form in the semi-supervised case and extends it to general blockwise-missingness settings.
\end{remark}

In what follows, we provide theoretical guarantees for the proposed estimator. For notational simplicity, we define $d(\widehat\boldeta,\boldeta^*)$ as a metric which measures the error of using $\widehat\boldeta$ to estimate the nuisance parameter $\boldeta$, whose formal definition is deferred to Section S.3.1 in the Supplementary Material. We also denote by $\|\bA\|_{\max}$ the elementwise maximum norm for a matrix $\bA$. 
\begin{theorem}\label{thm:main_nonsplit}
Suppose that  $\E_{P_0}(\|\widehat\bvphi(\bZ)-\bvphi(\bZ)\|_2^2)=o_P(1)$, $d(\widehat\boldeta^\init,\boldeta^*)=o_P(n^{-1/4})$, and $\|\widehat\bbeta-\bbeta^*\|_1+\|\widehat\var(\bbf)-\var_{P_0}(\bbf)\|_{\max}+\|\widetilde\bSigma_\ext-\bSigma_\ext\|_{\max}=o_P(1)$. Under certain regularity conditions (Assumptions S.1--S.4 in Section S.3.1), if $\E_{P_0}\{\tf_j(\bZ_{B_j})\}=\E_{Q_j}\{\tf_j(\bZ_{B_j})\}$,
% \ref{ass:donsker}--\ref{ass:vonMises} in Section \ref{sec:nonsplit_proof},
 we have
\[
n^{1/2}\bigl(\widehat{\btheta}^{\aos}-\btheta^*\bigr)\to_dN(0_d,\bSigma^*_{\tf}),
\]
where $\bSigma^*_{f}=\var_{P_0}(\bvphi)-\bbeta^{*\top}\{\bSigma_{\ext}+\var_{P_0}(\bbf)\}\bbeta^*$. Moreover, $\bSigma^*_{\bbf}$ can be consistently estimated by its sample version $\widehat\bSigma_{\tf}=n^{-1}\sum_{i\in S_0}\widehat\bvphi(\bZ_i)\{\widehat\bvphi(\bZ_i)\}^\top-\widehat\bbeta^\top\{\widehat\var(\bbf)+\widetilde\bSigma_\ext\}\widehat\bbeta$.
\end{theorem}

Valid Wald-type confidence intervals for $\btheta^*$ can be given using the asymptotic normality of $\widehat\btheta^\aos$ and consistency of $\widehat\bSigma_{f}$ to $\bSigma^*_{f}$. Theorem \ref{thm:main_nonsplit} shows that the proposed method enjoys the ``do-no-harm'' property in the sense that it is at least as efficient as the original one-step estimator, regardless of the sample size of external data and the quality of the chosen transfer function. Moreover, the proposed estimator enjoys a strict efficiency gain if the transfer function is informative in the sense that $\cov_{P_0}(\bbf,\bvphi)\neq \0$. The efficiency gain is more substantive for a larger external sample size: at the extreme case where $n/n_j\to0$ for all $j$, the efficiency gain is $\bbeta^{*\top}\var_{P_0}(\bbf)\bbeta^*=\cov_{P_0}(\bvphi,\bbf)\{\var_{P_0}(\bbf)\}^{-1}\cov_{P_0}(\bbf,\bvphi)$, which corresponds to the case as if $\E_{P_0}\{\bbf(\bZ)\}$ were known. Interestingly, the proposed estimator can deal with heterogeneous data distributions across
sites to some extent, e.g., $\var_{P_0}\{\tf_j(\bZ_{B_j})\}$ can be different from $\var_{Q_j}\{\tf_j(\bZ_{B_j})\}$, as long as the transfer function is chosen such that $\E_{P_0}\{\tf_j(\bZ_{B_j})\}=\E_{Q_j}\{\tf_j(\bZ_{B_j})\}$ holds.

The $n^{-1/4}$ rate of $\widehat\boldeta^\init$ and the consistency of $\widehat\bvphi(\bz)$ under the $L_2(P_0)$-norm are commonly required in the literature on one-step estimation and rate-doubly robust estimation \citep{smucler2019unifying,kennedy2022semiparametric}. Additionally, the condition that $\widehat\bbeta, \widehat\var(\bbf)$, and $\widetilde\bSigma_\ext$ are consistent is mild, allowing for an increasing number of external sites and high-dimensional transfer functions. For instance, suppose that $\tf_j(\bZ_{B_j})$ ($j=1,\dots,J$) are sub-Gaussian. Then, by Hoeffding's concentration inequality (Proposition 2.5 in \citealp{wainwright2019high}), we have
\[
\|\widehat\var(\bbf)-\var_{P_0}(\bbf)\|_{\max}+\|\widetilde\bSigma_\ext-\bSigma_\ext\|_{\max}\lesssim\bigl(n^{-1}\log p\bigr)^{1/2}+\min_{j}\bigl(n_j^{-1}\log p_j\bigr)^{1/2}.
\]
Moreover, under standard assumptions for high-dimensional linear regression (Theorem 7.13 of \citealp{wainwright2019high}), a lasso-type estimator satisfies $\|\widehat\bbeta-\bbeta^*\|_1\lesssim\|\bbeta^*\|_0(n^{-1}\log p)^{1/2}$ with high probability, where $\|\bbeta^*\|_0$ denotes the number of nonzero components of $\bbeta^*$. These results imply that the number of external sites or the data dimensionality can grow at a rate exponential in the internal sample size $n$. Therefore, the proposed method can be readily applied in multi-center research involving large-scale distributed systems and high-dimensional data.

It should be noted that the asymptotic variance of the augmented one-step estimator depends on the specific choice of the transfer function $\bbf$. In the next section, we will identify the optimal transfer function that gives rise to the most efficient augmented one-step estimator. Then, we propose using the ``kernel trick'' \citep{scholkopf2002learning} to estimate the optimal transfer function, with which we construct the data-driven augmented one-step estimator. We will show that under some conditions the data-driven augmented one-step estimator achieves the efficiency bound of a class of regular semiparametric models.

\subsection{Augmentation with data-driven transfer functions}\label{sec:aos_dd}
\subsubsection{Semiparametric efficiency bound for blockwise missing data}
\label{remark:mcar}
To identify the optimal transfer function and show the resulting estimator achieves the efficiency bound, we need first specify the underlying statistical model for the both internal and external data. Consider the random vector $\bR=(R_1,\dots,R_{J})$, where $R_j=1$ if $\bZ_{B_j}$ is observed and $R_j=0$ if $\bZ_{B_j}$ is missing. With blockwise missingness, either all blocks are observed or only one block is observed. Thus, the sample space of $\bR$ is $\{\be_1,\be_2,\dots,\be_J,\1_J\}$, where $\be_j$ denotes the $j$th standard basis vector in $\R^J$, and $\1_J$ is the $J$-dimensional vector of ones. 
% Specifically, $S_0=\{i:R_{i}=\1_J\}$ corresponds to the internal sample indices (where all blocks are observed), and $S_j=\{i:R_{i}=e_j\}$ corresponds to the indices of the $j$th external sample (where only the $j$th block is observed). 
To simplify the derivation of the semiparametric efficiency bound, we further impose the missing completely at random (MCAR) condition $\bR \ci \bZ$. We emphasize that this MCAR assumption is not required for Theorem \ref{thm:main_nonsplit} to hold; it is only used to derive the semiparametric efficiency bound in Proposition \ref{prop:semi_eff} below.

Define $N=\sum_{j=0}^J n_j$.
The observed data, including the internal sample $\{\bZ_i:i\in S_0\}$ and $J$ external samples $\{\bZ_{i,B_j}:i\in S_j\}$ ($j=1,\dots,J$), can be rewritten compactly as $\{(R_{i,1}\bZ_{i,B_1},\dots,R_{i,J}\bZ_{i,B_J})\}_{i=1}^N$, which are independent copies of $(R_1\bZ_{B_1},\dots,R_J\bZ_{B_J})$. Let $\widetilde{\cP}$ be the class of distributions for $(R_1\bZ_{B_1},\dots,R_J\bZ_{B_J})$ with restrictions 
\[
\bZ\sim P_0\in\cP,\quad \bR\ci \bZ,\quad0<\ve\le\min_j\P(\bR=\be_j)\le\max_j\P(\bR=\be_j)\le1-\ve<1
\]
for some constant $\ve$. Under the model $\widetilde{\cP}$, we can discuss semiparametric efficiency bound, and thus identify the optimal transfer function.

\begin{proposition}\label{prop:semi_eff}
Suppose that the observed data $\{(R_{i,1}\bZ_{i,B_1},\dots,R_{i,J}\bZ_{i,B_J})\}_{i=1}^N$ are independent samples drawn from some distribution $\widetilde{P}\in\widetilde\cP$. The function
\begin{equation*}
\widetilde\bvphi(\bZ,\bR;\btheta,\boldeta,\bbf^*)=\frac{I\bigl(\bR=\1_J\bigr)}{\P\bigl(\bR=\1_J\bigr)}\biggl\{\bvphi(\bZ;\btheta,\boldeta)-\sum_{j=1}^J\tf_j^*(\bZ_{B_j})\biggr\}+\sum_{j=1}^J\frac{I\bigl(\bR=\be_j\bigr)}{\P(\bR=\be_j)}\tf_j^*(\bZ_{B_j})
\end{equation*}
is the efficient influence function for $\btheta^*$, where $\tf_j^*$ ($j=1,\dots,J$) are the unique solution to the estimating equations
\begin{equation}\label{eq:choice_cf}
\E\biggl\{\bvphi(\bZ;\btheta^*,\boldeta^*)-\sum_{j=1}^J\gamma_{j,k}\tf_j^*(\bZ_{B_j})\mid \bZ_{B_k}\biggr\}=\0_d,\quad E\{\tf_k^*(\bZ_{B_k})\}=\0_d
\end{equation}
for $\gamma_{j,k}=1+I(j=k)\lim_nn/n_k$ with $j,k=1,\dots,J$.
Moreover, the semiparametric efficiency bound is $\var_{\widetilde{P}}\{\widetilde\bvphi(\bZ,\bR;\btheta^*,\boldeta^*,\bbf^*)\}$, which we denote by $\bV_\eff$.
\end{proposition}
The proof of Proposition \ref{prop:semi_eff} is provided in Section  S.3.2
% \ref{sec:aos_dd_proof}
of the Supplementary Material. The model class $\widetilde{\mathcal{P}}$ is somewhat restrictive, as it requires the observed variables at the external sites to share the same distribution as those in the internal site and assumes comparable sample sizes across sites. Nevertheless, such simplifications are standard in the literature for theoretical analysis. In particular, while not always necessary in practice, they are often imposed to establish statistical optimality—for instance, assuming identical covariate distributions and comparable sample sizes across labeled and unlabeled samples in semi-supervised learning frameworks \citep{zhang2019semi,hou2023surrogate}. This is exactly our case: the proposed AOS estimator continues to deliver efficiency gains even when the positivity assumption fails (i.e., $\mathbb{P}(\bR = \mathbf{1}_J) = 0$) and when data distributions are heterogeneous across sites (e.g. $\var_{P_0}\{\tf_j(\bZ_{B_j})\}\ne\var_{Q_j}\{\tf_j(\bZ_{B_j})\}$). Further details are provided in Theorem~\ref{thm:main_nonsplit} and the subsequent discussion.

Interestingly, Proposition~\ref{prop:semi_eff} encompasses existing results on identifying the optimal predictive models for PPI in semi-supervised learning. In particular, the optimal predictive model for PPI is any function $h^*(\bX)$ satisfying
\[
\E\{\nabla_{\btheta}\ell(\btheta^*;Y,\bX)\mid\bX\}=c(\nabla_{\btheta}\ell\{\btheta^*;h^*(\bX),\bX\}-\E[\nabla_{\btheta}\ell\{\btheta^*;h^*(\bX),\bX\}])
\] 
for some $c>0$ \citep[][Proposition 1]{angelopoulos2023anote}. By Remark~\ref{remark:ppi}, such a predictive model corresponds precisely to the optimal transfer function characterized in Proposition~\ref{prop:semi_eff}, up to a constant multiplicative factor. Thus, Proposition~\ref{prop:semi_eff} can be seen as a generalization of the PPI framework for identifying the optimal predictive model in semi-supervised learning to settings with blockwise missing data. 

\subsubsection{Data-driven augmented one-step estimator}
Suppose that we have obtained an estimator $\widehat{\bbf}=(\widehat\tf_1^\top,\dots,\widehat\tf_J^\top)^\top$ for the optimal transfer function $\bbf^*$, whose derivation will be discussed in the next subsection.
We can view $\widehat{\bbf}$ as a given transfer function. Applying \eqref{eq:betahat} gives the optimal transfer coefficients, and then invoking \eqref{eq:aos} yields the augmented one-step estimator. 
Specifically, let $\widehat{\tf}_{i,j}=\widehat{\tf}_j(\bZ_{i,B_j})$ be the evaluation of $\widehat{\tf}_j$ on $\bZ_{i,B_j}$ and $\widehat\bbf_i=(\widehat{\tf}_{i,1}^{\top},\dots,\widehat{\tf}_{i,J}^{\top})^\top$. 
The sample mean and variance of $\widehat{\tf}_j(\bZ_{B_j})$ for $j$th external site can be calculated, denoted by $\widetilde\bmu_{\ext,j}=n_j^{-1}\sum_{i\in S_j}\widehat{\tf}_{i,j}$ and $\widetilde\bSigma_{\ext,j}=n_j^{-1}\sum_{i\in S_j}\widehat{\tf}_{i,j}\widehat{\tf}_{i,j}^\top-\widetilde\bmu_{\ext,j}\widetilde\bmu_{\ext,j}^\top$, respectively. Thus, $\widetilde\bmu_\ext=(\widetilde\bmu_{\ext,1}^\top,\dots,\widetilde\bmu_{\ext,J}^\top)^\top$ and $\widetilde\bSigma_\ext=\text{diag}(n\widetilde\bSigma_{\ext,1}/n_1,\dots,n\widetilde\bSigma_{\ext,J}/n_J)$ are estimates for the mean and variance of $\widehat{\bbf}$ at external sites. The sample variance of $\widehat{\bbf}$ and the covariance between $\widehat\bbf$ and $\widehat\bvphi$ can also be computed in the internal site, denoted by $\widehat\var(\widehat\bbf)$  and $\widehat\cov(\widehat\bbf,\widehat\bvphi)$, respectively. Invoking \eqref{eq:betahat}, we can obtain $\widehat\bbeta_\dd$ as an estimate for the optimal transfer coefficients. By \eqref{eq:aos}, our data-driven augmented one-step estimator is
\begin{equation}\label{eq:aos_dd}
\widehat\btheta^\aos_\dd=\widehat\btheta^\init+n^{-1}\sum_{i\in S_0}\bigl\{\bvphi(\bZ_i;\widehat\btheta^\init,\widehat\boldeta^\init)-\widehat\bbeta_\dd^\top\bigl(\widehat\bbf_i-\widetilde\bmu_\ext\bigr)\bigr\}.
\end{equation}

In what follows, we investigate the theoretical properties of the data-driven augmented one-step estimator $\widehat\theta^\aos_\dd$. A key observation is that statistical validity of $\widehat\theta^\aos_\dd$ does not require the working model for estimating the optimal transfer function to be correctly specified.
Let $\widetilde{\tf}_{j}^*(\bZ_{B_j})$ denote the probabilistic limit of $\widehat{\tf}_j(\bZ_{B_j})$ in the sense that $\E_{P_0}\{\sum_{j=1}^J\|\widehat{\tf}_j(\bZ_{B_j})-\widetilde{\tf}_j^*(\bZ_{B_j})\|^2\}=o(1)$. Write $\widetilde\bbf^*(\bZ)=(\widetilde{\tf}_{1}^{*\top}(\bZ_{B_1}),\dots,\widetilde{\tf}_{J}^{*\top}(\bZ_{B_J}))^\top$. Continuous mapping theorem suggests that $\widehat\bbeta_\dd$ should be consistent for
\[
\bbeta^*_\dd\equiv\{\var_{P_0}(\widetilde\bbf^*)+\Sigma_\ext^*\}^{-1}\cov(\widetilde\bbf^*,\bvphi),
\]
where $\bSigma_\ext^*=\text{diag}\{\rho_1\var_{Q_1}(\widetilde{\tf}_1^*(\bZ_{B_1})),\dots,\rho_J\var_{Q_J}(\widetilde{\tf}_J^*(\bZ_{B_J}))\}$ is the probabilistic limit of $\widetilde\bSigma_\ext$ and $\rho_j=\lim_nn/n_j$. It is then expected that the AOS estimator using $\widehat\bbf$ as the aggregated transfer function is asymptotically equivalent to that using $\widetilde\bbf^*$. Consequently, even if the working model for estimating the optimal transfer function is misspecified, efficiency gains may still be achieved, albeit not the maximum possible gain. The main theoretical properties of $\widehat\theta^\aos_\dd$ are summarized below.
\begin{theorem}\label{thm:aos_dd} Under all the conditions required in Theorem \ref{thm:main_nonsplit} and Assumption S.5
in Section S.3.2,
% Assumption \ref{ass:donsker_cf} in Section \ref{sec:aos_dd_proof}, 
we have
\[
n^{1/2}\bigl(\widehat\btheta^\aos_\dd-\btheta^*\bigr)\to_dN(\0,\bSigma_{\widetilde\bbf^*}^*),
\]
where $\bSigma_{\widetilde\bbf^*}^*=\var_{{P_0}}\{\bvphi(\bZ;\btheta^*,\boldeta^*)\}-\bbeta_\dd^{*\top}\{\var_{{P_0}}(\widetilde\bbf^*)+\bSigma_\ext^*\}\bbeta_\dd^*$ can be consistently estimated by
$\widehat\bSigma_{\widehat\bbf}=n_0^{-1}\sum_{i\in S_0}\widehat\bvphi(\bZ_i)\{\widehat\bvphi(\bZ_i)\}^\top-\widehat\bbeta_\dd^\top\bigl\{\widehat\var(\widehat\bbf)+\widetilde\bSigma_\ext\bigr\}\widehat\bbeta_\dd$.
Moreover, if $\bbf^*=\widetilde{\bbf}^*$, then $\widehat\btheta^\aos_\dd$ achieves the semiparametric efficiency bound $\bV_\eff$.
\end{theorem}
Valid Wald-type confidence intervals for $\btheta^*$ can be provided using the asymptotic normality of $\widehat\btheta^\aos_\dd$ and consistency of $\widehat\bSigma_{\widehat\bbf}$ to $\bSigma_{\widetilde\bbf^*}^*$. Assumption S.5
% \ref{ass:donsker_cf} 
restricts the model complexity for estimating the optimal transfer function. Theorem \ref{thm:aos_dd} shows that the data-driven augmented one-step estimator also enjoys the ``do-no-harm'' property. Furthermore, when the working model for the transfer function is correctly specified, the resulting estimator achieves optimality in the local minimax sense.

% In general, the optimal transfer function that minimizes the asymptotic variance does not have a closed-form solution. However, it can be explicitly characterized in certain special cases, as illustrated below.
% \begin{example}[Semi-supervised learning]
% Consider a setting where the internal study observes $Z=(Z_{B_1}^\top,Z_{B_1^c}^\top)^\top$ while the external site can only observe $Z_{B_1}$. In this scenario, equation \eqref{eq:choice_cf} simplifies to
% \[
% f_1^*(Z_{B_1})=\P(R=e_1)\E\{\bvphi(Z;\btheta^*,\boldeta^*)\mid Z_{B_1}\}.
% \]
% To estimate the optimal transfer function, it therefore suffices to estimate $\E\{\bvphi(Z;\btheta^*,\boldeta^*)\mid Z_{B_1}\}$.
% \end{example}
% \begin{example}[Independent blocks] Now consider a case where the data blocks are pairwise independent, i.e., $Z_{B_i}\ci Z_{B_j}$ for $1\le i\ne j\le J$. Under this assumption, \eqref{eq:choice_cf} reduces to
% \[
% f_k^*(Z_{B_k})=\frac{\P(R=e_k)}{\P(R=e_k)+\P(R=\1_J)}\E\{\bvphi(Z;\btheta^*,\boldeta^*)\mid Z_{B_k}\},\quad\text{for }k=1,\dots,J.
% \]
% As in the previous example, estimating the optimal transfer function in this setting requires only the estimation of $\E\{\bvphi(Z;\btheta^*,\boldeta^*)\mid Z_{B_k}\}$ for each $k$.
% \end{example}
\subsubsection{Estimation of the optimal transfer function}
In this subsection, we consider the estimation of the optimal transfer function $\bbf^*$. Equations in \eqref{eq:choice_cf} are essentially conditional
moment restrictions \citep{NEWEY1993419}, and we can use the ``kernel trick'' to solve them. For simplicity, suppose that $d=1$ and $\gamma_{j,k}$ are known. We write $\tf_j$ as $f_j$ in this section since the optimal transfer function is a scalar. 
By the law of iterated expectation and the definition of conditional expectation, \eqref{eq:choice_cf} is equivalent to
\[
\E\biggl[\biggl\{\varphi(\bZ)-\sum_{j=1}^J\gamma_{j,k}f_j^*(\bZ_{B_j})\biggr\}g_k(\bZ_{B_k})\biggr]=0,\quad (k=1,\dots,J)
\]
for all measurable function $g_k:\bZ_{B_k}\mapsto g_k(\bZ_{B_k})\in\R$.
Thus, to estimate $f_j^*$ we can minimize the risk function against the worst-case choice of $g_1,\dots,g_J$:
\[
R(f_1,\dots,f_J)=\sup_{g_1,\dots,g_J}\sum_{k=1}^J\biggl(\E\biggl[\biggl\{\varphi(\bZ)-\sum_{j=1}^J\gamma_{j,k}f_j(\bZ_{B_j})\biggr\}g_k(\bZ_{B_k})\biggr]\biggr)^2.
\]
If $g_k$ is an element of a reproducing kernel Hilbert space, i.e., $g_k\in\cH_k\subset L_2(Q_k)$, whose associated kernel function is $H_k(\bZ_{B_k},\bZ_{B_k}')$, then $R(f_1,\dots,f_J)$ can be rewritten in the form
\[
\sum_{k=1}^J\E\biggl[\biggl\{\varphi(\bZ)-\sum_{j=1}^J\gamma_{j,k}f_j(\bZ_{B_j})\biggr\}H_k(\bZ_{B_k},\bZ_{B_k}')\biggl\{\varphi(\bZ')-\sum_{j=1}^J\gamma_{j,k}f_j(\bZ_{B_j}')\biggr\}\biggr].
\]
Moreover, under additional assumptions, $R(f_1,\dots,f_J)=0$ if and only if they satisfy \eqref{eq:choice_cf}. We refer to Lemma S.5 in Section S.3.2
%  Lemma \ref{lem:U_stat} in Section \ref{sec:aos_dd_proof} 
for a formal statement.  The ``kernel trick'' have been applied to nonparametric estimation of causal effects in settings with instruments and negative control variables \citep{mastouri2021proximal,zhang2023instrumental}. In this work, we extend these techniques to accommodate multiple, overlapping conditional moment restrictions, thereby generalizing the existing methodology to more complex scenarios.

In practice, to find estimators for $\bbf^*=(f_1^*,\dots,f_J^*)^\top$, one can minimize the empirical version of $R(f_1,\dots,f_J)$ over a suitably chosen class of working models $\cC_j$, and we denote by $\widehat{\bbf}=(\hat{f}_1,\dots,\hat{f}_J)$ the resultant estimator for $\bbf^*$. We give two illustrative examples as follows.
\begin{example}[Pairwise independent blocks]\label{example:pair} Suppose that $\bZ_{B_i}\ci \bZ_{B_j}$ for $i\ne j$. Lemma S.6 % \ref{lem:pair_ind} 
 shows that minimizing $R(f_1,\dots,f_J)$ is separable with respect to each $f_j$; that is, we need only consider
\[
R(f_j)=\E\bigl[\bigl\{\varphi(\bZ;\btheta^*,\boldeta^*)-\gamma_{k,k}f_k(\bZ_{B_k})\bigr\}H_k(\bZ_{B_k},\bZ_{B_k'})\bigl\{\varphi(\bZ';\btheta^*,\boldeta^*)-\gamma_{k,k}f_k(\bZ_{B_k}')\bigr\}\bigr].
\]
When $H_k(\bZ_{B_k},\bZ_{B_k}')=I(\bZ_{B_k}=\bZ_{B_k'})$, $R(f_j)$  reduces to $\E\bigl[\bigl\{\varphi(\bZ;\btheta^*,\boldeta^*)-\gamma_{k,k}f_k(\bZ_{B_k})\bigr\}^2\bigr]$, whose empirical version corresponds to the ordinary least square estimation for $f_k^*$.
% $\E\{\bvphi(Z)\mid Z_{B_1},\dots,Z_{B_J}\}=g_1^*(Z_{B_1})+\dots+g_J^*(\bZ_{B_J})$ for some measurable functions $g_1^*,\dots,g_J^*$.
\end{example}
\begin{example}[Linear models]\label{example:mr_linear}
Let $f_j(\bZ_{B_j})=\ba_j^\top\bb_j(\bZ_{B_j})$ with $\bb_j=(b_{j,1},\dots,b_{j,q_j})^\top$ being a vector of known basis functions, $b_{j,1}=1$ the intercept, and $\ba_j$ the coefficient vector. Define $\balpha=(\ba_1^\top,\dots,\ba_J^\top)^\top\in\R^q$ with $q=\sum_{j=1}^Jq_j$ and $\bB(\bZ)=\mathrm{blockdiag}\{\bb_1(\bZ_{B_1}),\dots,\bb_J(\bZ_{B_J})\}\in\R^{q\times J}$. Minimizing the empirical version of $R(f_1,\dots,f_J)$ with respect to $f_j$ simplifies to the following convex optimization problem:
\[
\min_{\balpha\in\R^q}\sum_{i\ne j\in S_0}\Bigl\{\balpha^\top\bB(\bZ_i)\bGamma\widetilde{\bH}(\bZ_i,\bZ_j)\bGamma^\top\{\bB(\bZ_j)\}^\top\balpha-2\balpha^\top\bB(\bZ_i)\bGamma\widetilde{\bH}(\bZ_i,\bZ_j)\1_J\widehat\varphi(\bZ_j)\Bigr\}+\lambda J(\balpha),
\]
where $\bGamma$ denotes the matrix whose $(j,k)$th entry is $\gamma_{j,k}$, $\widetilde{\bH}(\bZ,\bZ')$ is a diagonal matrix with $k$th diagonal is $H_k(\bZ_{B_k},\bZ_{B_k}')$,  $J(\balpha)$ is the regularization term, and $\lambda$ is the tuning parameter. In fact, when we choose the ridge regularization, i.e., $J(\balpha)=\|\balpha\|_2^2$, the global minimizer of the objective function, denoted by  $\widehat\balpha=(\widehat\ba_1^\top,\dots,\widehat\ba_J^\top)^\top$, has an explicit form:
\[
\widehat\balpha=\biggl\{\sum_{i\ne j\in S_0}\bB(\bZ_i)\bGamma\widetilde{\bH}(\bZ_i,\bZ_j)\bGamma^\top\{\bB(\bZ_j)\}^\top+\lambda I_{q\times q}\biggr\}^{-1}\biggl\{\sum_{i\ne j\in S_0}\bB(\bZ_i)\bGamma\widetilde{\bH}(\bZ_i,\bZ_j)\1_J\widehat\varphi(\bZ_j)\biggr\}.
\]
The estimate for the transfer function $f_j^*(\bZ_{B_j})$ is $\widehat{f}_j(\bZ_{B_j})=\widehat\ba_j^\top\bb_j(\bZ_{B_j})$, $j=1,\dots,J$.
The linear model $f_j(\bZ_{B_j})=\ba_j^\top\bb_j(\bZ_{B_j})$ can be extended to general kernel-based methods \citep{scholkopf2000kernel}.
\end{example}

\subsection{Communication cost for distributed blockwise missing data}
Reducing the number of communication rounds is crucial for large-scale networks \citep{li2024centralized}, since frequent updates can be time-consuming and often require a sophisticated automated infrastructure \citep{wu2012g}.
One-shot algorithms provide a promising solution to these issues because they need only a single round of cross-site data exchange \citep{li2024centralized}. Such a setup not only lowers the burden on the research network’s infrastructure but also simplifies manual handling of data. By avoiding repeated exchanges, one-shot approaches significantly reduce the risk of privacy exposure and minimize the coordination efforts required among data partners, thus promoting scalability. 
% Furthermore, these algorithms sidestep the synchronization bottlenecks common to iterative methods: when iterative updates depend on all sites submitting new summaries in each round, even small delays can hold back the entire process. In contrast, one-shot estimators, such as the methods proposed by \citet{luo2022dlmm}, can be calculated as soon as each site provides its local summary data, thereby removing synchronization barriers and further enhancing scalability. In sum, to effectively handle blockwise missingness in distributed networks, it is critical to develop algorithms, especially one-shot methods, that are both communication-efficient and statistically sound.

It is worth mentioning that the (data-driven) augmented one-step estimator requires only a single round-trip between internal and external sites. Specifically, the internal site first sends the estimated transfer function $\widehat{\tf}_j$ (for example, $\widehat{\ba}_j$ in Example \ref{example:mr_linear}) to the external sites and then requests summary-level statistics $(n_j, \widetilde{\bmu}_{\text{ext}}, \widetilde{\bSigma}_{\text{ext}})$ from them. Then, applying \eqref{eq:betahat} at the internal site we can obtain the estimate for the optimal transfer coefficients. Invoking \eqref{eq:aos} or \eqref{eq:aos_dd} yields the final augmented one-step estimator. This single round-trip communication protocol is efficient and practical for large-scale distributed networks, where minimizing data transfer is essential for both speed and security.
\section{Simulation}
\label{sec:sim}
We compare four estimators: an initial estimator derived solely from the internal sample, an augmented one-step estimator with $\lambda=0$ in \eqref{eq:betahat}, an augmented one-step estimator with $J(\bbeta)=\|\bbeta\|_1$ in \eqref{eq:betahat} for adaptively selecting external sites, and a data-driven augmented one-step estimator. We term them Naive, AOS, AOS-Lasso, and DD-AOS respectively. The performance metrics include the absolute relative bias (ARB, the absolute value of the bias of an estimator divided by the true value of the parameter), the mean squared error (MSE), and the empirical variance (EV). 
We also examine the relative efficiency gain (REG), defined as one minus the ratio of the empirical variance of an estimator to that of the initial estimator. In addition, we report the coverage probability (CP) and average length (AL) of the associated 95\% Wald-type confidence intervals. 

We conduct simulations to evaluate the performance from three perspectives: the efficiency gain compared to the initial one-step estimator, the robustness to an increasing number of uninformative external sites, and the additional efficiency gain provided by a data-driven transfer function. Three distinct scenarios are designed to examine each of these aspects. 
% Due to space limitations, we only report the results for the third simulation design. 
% Additional results are deferred to Section \ref{sec:additional_sim} in the Supplementary Material.

In the first simulation design, our target parameter is the expectation of the outcome $Y$. The data generating model is 
\begin{equation}
Y=2+X_1+2X_2+3X_3+\{15(\rho^{-2}-1)\}^{1/2}\epsilon,
\label{eq:Y_sim}
\end{equation}
where $\epsilon\sim N(0,1)$ denotes the additive noise, $\bX=(X_1, X_2, X_3)^\top$ follows $N(0, I_3)$, and $\rho$ is the correlation between $\E(Y\mid \bX)$ and $Y$.
Under this model, the true value of $\E(Y)$ is $\theta^*=2$.
The distributed system includes one internal site and $J=50$ external sites.
At the internal site, the data $\{\bZ_i=(Y_i, \bX_i^\top)^\top\}_{i=1}^n$ is fully observed.
At each external site, only one of the three covariates ($X_1$, $X_2$, or $X_3$) is observed, with sample-size $n_j$.
We use the identity transfer function; that is, $\bbf(\bZ)=(Z_{B_1}, \dots, Z_{B_{50}})^\top$, where $Z_{B_j}$ is the observed covariate at the $j$th external site.

By Theorem \ref{thm:main_nonsplit}, if $\E(\bX)$ were known, the best possible relative efficiency gain would be $\var\{\E(Y\mid \bX)\}/\var(Y)=\rho^2$.  In practice, $\E(\bX)$ is not known, so the attainable gain is inevitably smaller. By varying $\rho$, we can examine how well our proposed method approaches this theoretical upper bound.
\begin{table}[htp]
\centering
\caption{{Performance of the proposed methods and the naive sample mean under correlations $\rho=0.3,0.5$, and $0.7$ and sample sizes $n=n_j=100$. All values have been multiplied by $10^2$.}}
\resizebox{0.6\textwidth}{!}{ 
\begin{tabular}{c c c c c c c c}
\(\rho\) & Method &ARB & MSE & EV  &REG  & \%CP & AL \\[0.5ex]

0.3 & Naive       & 0.9 & 149.4 & 149.3 & 0.0 & 94.5 & 486.7 \\
& AOS         & 0.9 & 143.2 & 143.2 & 4.1 & 94.2 & 458.1 \\
& AOS-Lasso   & 0.2 & 141.9 & 142.9& 5.0 & 94.5 & 465.9 \\[1ex]
0.5 & Naive       & 0.6 & 53.6  & 53.6& 0.0  & 95.1 & 292.3 \\
& AOS         & 0.6 & 43.1  & 43.1 &19.7  & 94.1 & 251.3 \\
& AOS-Lasso   & 0.0 & 42.6  & 42.6 &20.5 & 94.5 & 255.8 \\[1ex]
0.7 & Naive       & 0.4 & 27.2  & 27.2&0.0  & 95.7 & 208.9 \\
& AOS         & 0.4 & 15.5  & 15.5 &43.2 & 94.0 & 150.7 \\
& AOS-Lasso   & 0.1 & 15.3  & 15.3 &43.8  & 94.2 & 153.0 \\[1ex]
\end{tabular}
}
\label{table:n=100,rho}
\end{table}

Table \ref{table:n=100,rho} summarizes the performance of the proposed estimators and the naive sample mean under correlations $\rho=0.3,0.5,0.7$ when $n=n_j=100$. Based on 1,000 replicates, the AOS estimators consistently yield smaller mean squared errors and empirical variances, produce narrower confidence intervals with comparable coverage probabilities, and exhibit greater efficiency gains at higher correlations. Additional results for 
$n=50, n_j=100$ (Table \ref{table:n=50,rho}) confirm these findings.

The second simulation design examines a high-dimensional setting to assess how the proposed methods perform when many external sites offer little useful information. The outcome $Y$ follows the same model as \eqref{eq:Y_sim} with $\rho=0.5$, but we now have 50-dimensional covariates $\bX=(X_1, \dots, X_{50})^\top$, where each $X_j\sim N(0,1)$ independently.
The target parameter is still $\theta=\E(Y)$ with $\theta^*=2$.
Data are fully observed at the internal site, whereas at the $j$th external site, only $(X_{5(j-1)+1}, \dots, X_{5j})$ is observed for $j=1, \dots, 10$.
Except for the first external site, the remaining external sites provide no additional information on $\theta$ due to the independence of covariates.
\begin{table}[ht]
\centering
\caption{{Performance of the proposed methods and the naive sample mean, under correlations \(\rho=0.3\), \(0.5\), and \(0.7\) and sample sizes \(n=50\) and \(n_j=100\). All values have been multiplied by $10^2$.}}
\resizebox{0.6\textwidth}{!}{ 
\begin{tabular}{c c c c c c c c}
\(\rho\) & Method      &ARB   & MSE & EV & REG & \%CP & AL \\[0.5ex]
0.3      & Naive       & 2.1 & 312.7 & 312.6 & 0.0 & 94.4 & 682.5 \\
& AOS         & 1.2 & 299.9 & 299.8 & 4.1 & 92.3 & 632.1 \\
& AOS-Lasso   & 1.3 & 295.8 & 295.7 &5.4 & 93.5 & 653.1 \\[1ex]
0.5      & Naive       & 1.1 & 111.1 & 111.0 & 0.0 & 94.0 & 409.7 \\
& AOS         & 1.1 & 89.0  & 88.9 & 19.9  & 92.2 & 345.6 \\
& AOS-Lasso   & 0.0 & 88.7 & 88.7 & 20.1 & 93.6 & 362.0 \\[1ex]
0.7      & Naive       & 0.6 & 55.7 & 55.7 &0.0 & 93.5 & 292.8 \\
& AOS         & 0.7 & 31.1  & 31.1 & 44.1& 92.1 & 205.6 \\
& AOS-Lasso   & 0.2 & 31.2 & 31.2& 44.0 & 93.7 & 216.3 \\[1ex]
\end{tabular}
}
\label{table:n=50,rho}
\end{table}

\begin{figure}[htp]
\centering
\includegraphics[width=0.8\linewidth]{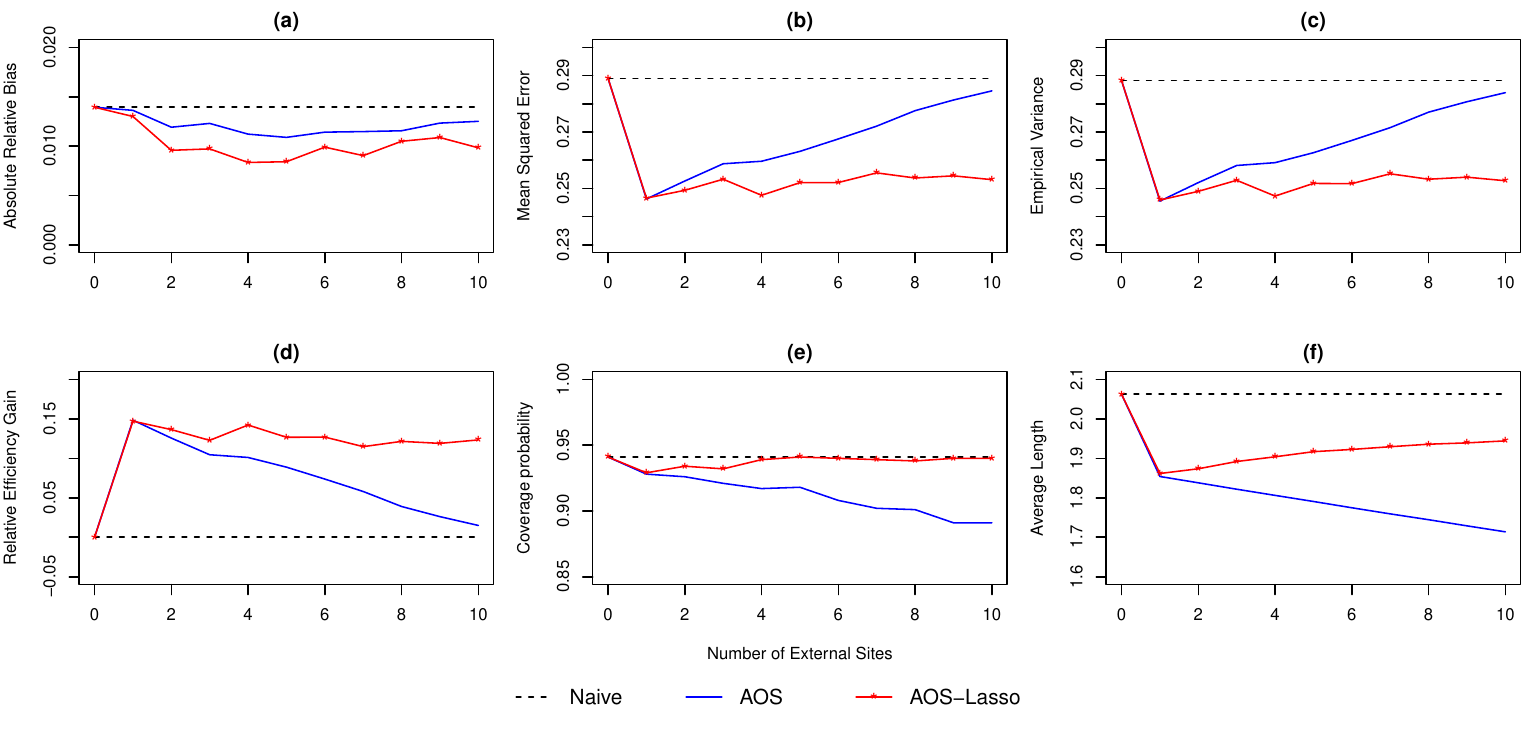}
\caption{{Performance of the proposed methods and the naive sample mean as the number of external sites increases with sample sizes $n=200$ and $n_j=500$, evaluated with respect to: (a) absolute relative bias; (b) mean squared error; (c) empirical variance; (d) relative efficiency gain; (e) coverage probability; (f) average length.}}
\label{fig:n_j=500}
\end{figure}

\begin{figure}[htp]
\centering
\includegraphics[width=0.9\linewidth]{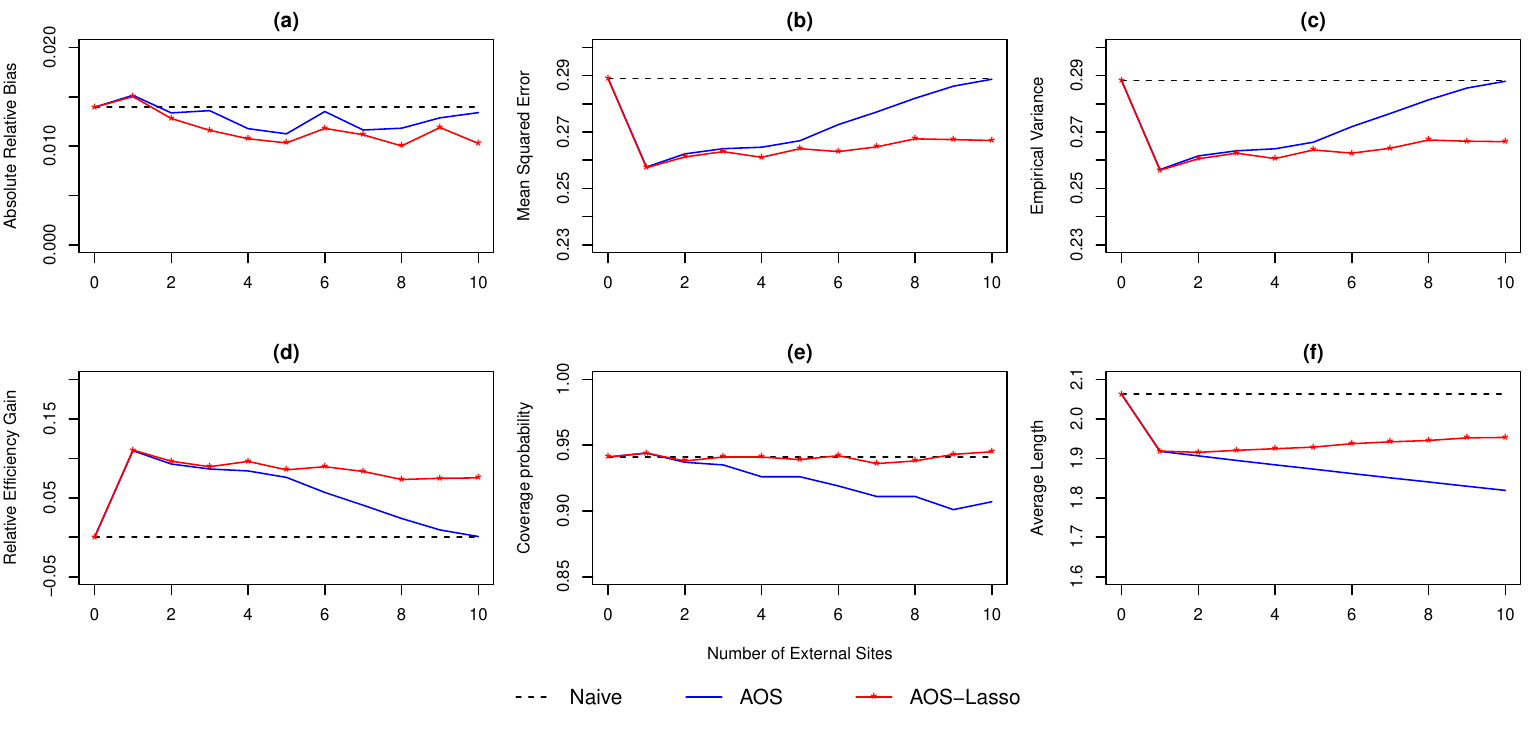}
\caption{{Performance of the proposed methods and the naive sample mean as the number of external sites increases, when $n=n_j=200$, evaluated with respect to: (a) absolute relative bias; (b) mean squared error; (c) empirical variance; (d) relative efficiency gain; (e) coverage probability; (f) average length.}}
\label{fig:n_j=200}
\end{figure}

\begin{figure}[htp]
\centering
\includegraphics[width=0.9\linewidth]{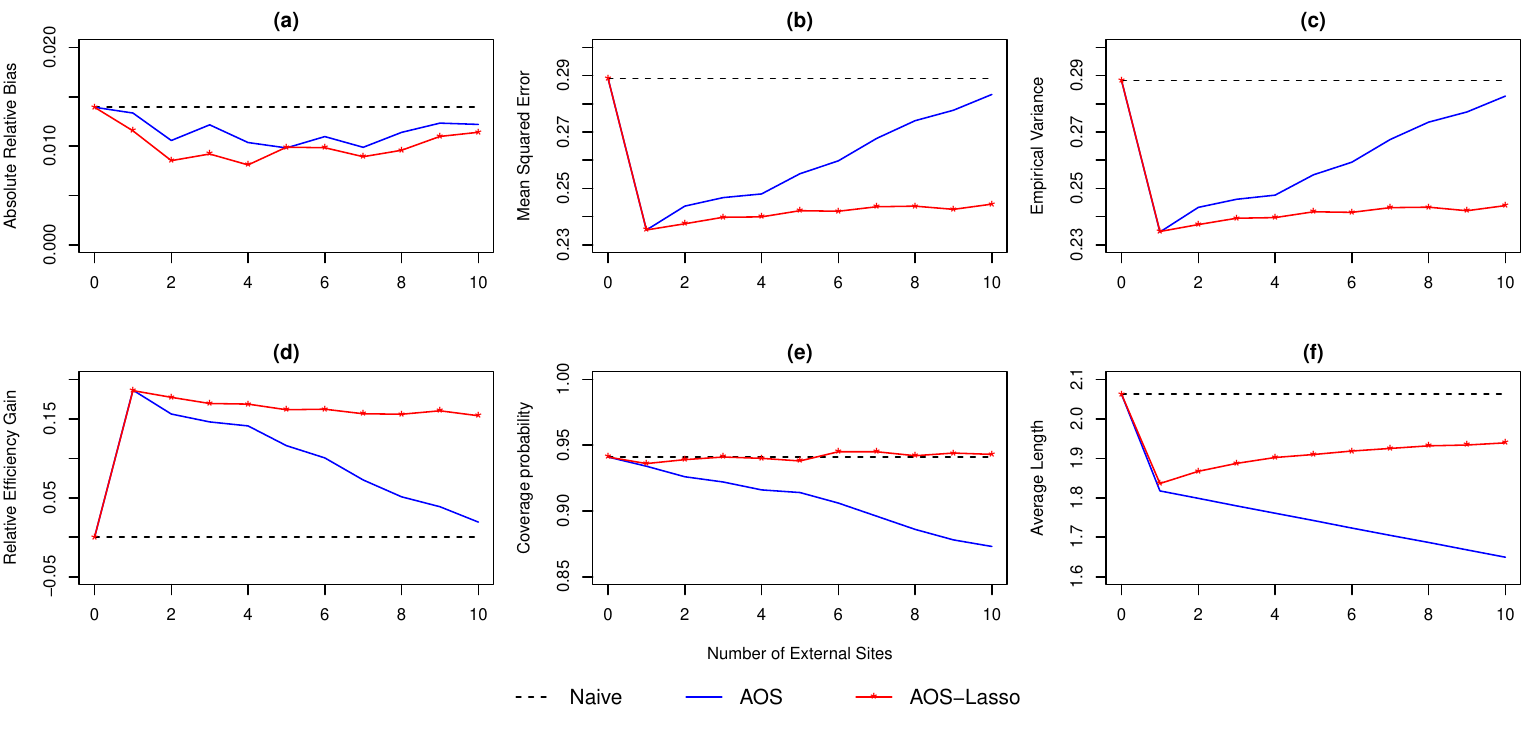}
\caption{{Performance of the proposed methods and the naive sample mean as the number of external sites increases, when $n=200$ and $n_j=1000$, evaluated with respect to: (a) absolute relative bias; (b) mean squared error; (c) empirical variance; (d) relative efficiency gain; (e) coverage probability; (f) average length.}}
\label{fig:n_j=1000}
\end{figure}

Figure \ref{fig:n_j=500}, for sample sizes $n=200$ and $n_j=500$, indicates that the AOS estimators outperform the Naive estimator by achieving lower mean squared error and empirical variance.
However, as the number of uninformative external sites increases, the performance of the AOS estimator converges towards that of the Naive estimator, while the AOS-Lasso estimator remains relatively unaffected.
This highlights the robustness of the regularized AOS estimator to a growing number of uninformative external sites.
In terms of confidence intervals, the AOS-Lasso estimator maintains coverage probabilities close to the nominal $95\%$ level with a shorter average length than the Naive estimator, whereas the the AOS estimator fails.
Figures \ref{fig:n_j=200} and \ref{fig:n_j=1000} present simulation results under the same setting but with different external sample sizes ($n_j=200$ and $n_j=1000$, respectively). 
These results are consistent with the observations described above.
A comparison across the three figures reveals that larger external sample sizes lead to greater efficiency gains.

In the third simulation, we assess the additional efficiency gains provided by data-driven transfer functions. The data generating model is
\[
Y=X_1+X_2+X_3+2X_1X_2+2X_2X_3+2X_1X_3+0.5\ve,
\]
where $\bX=(X_1,X_2,X_3)^\top\sim N(0,\bSigma)$ and $\ve\sim N(0,1)$. The covariance matrix is $\bSigma=\varrho\1_3\1_3^\top+(1-\varrho)\bI_3$, where $\varrho$ is the pairwise correlation between covariates. Under this model, the true value of $\E(Y)$ is $\theta^*=6\varrho$. The distributed system includes one internal site and $J=3$ external sites.
At the internal site, the data $\{\bZ_i=(Y_i, \bX_i^\top)^\top\}_{i=1}^n$ is fully observed, and each external site $j$ ($j=1,2,3$) only observes $X_j$ with a sample size of $n_j$.
We follow Example \ref{example:mr_linear} to obtain estimators for the transfer function. Specifically, we use the basis functions $\bb_j(\bZ)=(1,X_j,X_j^2)^\top$ and set the regularization parameter $\lambda=0$. 

According to Example \ref{example:pair}, when covariates are pairwise independent, it is sufficient to estimate the transfer function for each external site separately. The optimal transfer function in the case of $\varrho=0$ is $f_j^*(X_j)=\E(Y\mid X_j)-\E(Y)=X_j-\E(Y)$, which is the identity function with a constant shift. By varying the correlation level $\varrho$, we can compare DD-AOS with AOS, where the former is obtained by jointly estimating transfer functions using the method outlined in Example \ref{example:mr_linear} and the latter obtained using the identity transfer function. 

Table \ref{table:n=100,varrho} summarizes the results.  Based on 1,000 replicates, the AOS and DD-AOS estimators consistently achieve lower mean squared errors and empirical variances compared to the naive method. They also produce narrower confidence intervals while maintaining comparable coverage probabilities. Overall, DD-AOS outperforms the Naive and AOS methods, particularly in settings with elevated pairwise correlations, offering more efficient and accurate estimations.
Interestingly, under small pairwise correlation ($\varrho=0.1$), AOS outperforms DD-AOS, possibly because in this case the identity function approximates the optimal transfer function better than the data-driven transfer function does. As the correlation $\varrho$ increases, AOS has a decreased relative efficiency gain, while DD-AOS achieves higher relative efficiency gain values. This reflects its enhanced performance in leveraging the correlated data and highlights the potential applicability of the proposed data-driven augmented one-step estimator in analyzing multi-modal data, which is known for high correlation among modalities \citep{lahat2015multimodal}.
\begin{table}[htbp]
\centering
\caption{{Performance of the proposed methods and the naive sample mean under pairwise correlations $\varrho=0.3,0.5$, and $0.7$ and sample sizes $n=n_j=200$. All values have been multiplied by $10^2$.}}
\resizebox{0.6\textwidth}{!}{ 
\begin{tabular}{c c c c c c c c}
\(\varrho\) & Method & ARB& MSE & EV &REG& \%CP & AL\\[0.5ex]
0.1 & Naive       & 0.5 & 9.3 & 9.3 & 0.0 & 94.8 & 118.1 \\
& AOS         & 0.8 & 8.2 & 8.2 &11.8 & 94.5 & 110.9 \\
& DD-AOS   & 1.8 & 8.6 & 8.6 & 7.5 & 94.4 & 112.1 \\[1ex]
0.3 & Naive       & 0.1 & 13.9 & 13.9 & 0.0 & 93.8 & 143.3 \\
& AOS         & 0.7 & 12.6 & 12.6 & 9.4& 93.7 & 134.3 \\
& DD-AOS   & 1.9 & 10.4 & 10.3 & 25.9 & 94.7 & 123.5 \\[1ex]
0.5 & Naive       & 0.2 & 19.7  & 19.7& 0.0  & 93.7 & 171.5 \\
& AOS         & 0.5 & 18.0 & 18.0 &8.6 & 93.5 & 160.9 \\
& DD-AOS   & 1.6 & 11.2  & 10.9 &44.6& 94.0 & 126.6 \\[1ex]
0.7 & Naive& 0.2&27.4 & 27.4&0.0& 93.7&200.8 \\
& AOS      & 0.6& 25.5& 25.4& 7.3  & 93.5&189.1  \\
& DD-AOS   & 1.8& 11.2& 10.7& 61.0  &93.4 & 125.2 \\[1ex]
% 0.9 & Naive       & 0.1 &36.7 & 36.7 & 0.0 & 93.3 & 230.6 \\
% & AOS         & 0.6& 33.8 & 33.7 & 8.1& 93.3 & 218.5 \\
% & DD-AOS   & 1.6 & 11.2 & 10.4 & 71.7 & 91.7 & 122.8 \\[1ex]
\end{tabular}
}
\label{table:n=100,varrho}
\end{table}

% \section{Real-world data analysis}

\section{Discussion}
This paper introduces a comprehensive framework for distributed inference in the presence of blockwise missingness, building on one-step estimators. By effectively utilizing external blockwise missing data through transfer functions, the proposed estimator requires only one-shot communication and enjoys the ``do-no-harm'' property. We also consider two extensions, the regularized and the data-driven augmented one-step estimators. The regularized estimator presents notable efficiency gains consistently as the number of external sites grows. The data-driven estimator attains the semiparametric efficiency bound and demonstrates substantial robustness to highly correlated blocks of covariates, thereby establishing a theoretical foundation for its potential in analyzing multimodal data. Moreover, the proposed method preserves the broad applicability of traditional one-step estimators, encompassing tasks such as causal effect estimation, high-dimensional hypothesis testing, and variable importance assessment \citep{fisher2021visually,dukes2024doubly,ning2017general,williamson2023general,wolock2023nonparametric}. This wide-ranging applicability warrants further exploration in practical applications.

% \section{Disclosure statement}\label{disclosure-statement}

% The authors have the following conflicts of interest to declare (or
% replace with a statement that no conflicts of interest exist).

% \section{Data Availability Statement}\label{data-availability-statement}

% Deidentified data have been made available at the following URL: XX.

% \phantomsection\label{supplementary-material}
% \bigskip

% \begin{center}

% {\large\bf SUPPLEMENTARY MATERIAL}

% \end{center}

% \begin{description}
% \item[Title:]
% Brief description. (file type)
% \item[R-package for MYNEW routine:]
% R-package MYNEW containing code to perform the diagnostic methods
% described in the article. The package also contains all datasets used as
% examples in the article. (GNU zipped tar file)
% \item[HIV data set:]

% \end{description}

% \section{BibTeX}\label{bibtex}

% We encourage you to use BibTeX. If you have, please feel free to use the
% package natbib with any bibliography style you're comfortable with. The
% .bst file agsm has been included here for your convenience. 

\bibliographystyle{apalike}
\bibliography{bibliography}

\end{document}